\documentclass[conference]{IEEEtran}
\IEEEoverridecommandlockouts
\usepackage{cite}
\usepackage{amsmath,amssymb,amsfonts}
\usepackage{algorithmic}
\usepackage{graphicx}
\usepackage{float}
\usepackage{textcomp}
\usepackage{xcolor}
\usepackage{subfigure}
\usepackage{url}
\usepackage{algorithm}
\usepackage{algorithmic}
\usepackage{makecell}
\usepackage{color}
\usepackage{verbatim}
\usepackage{hyperref}
\usepackage{fancyhdr}
\usepackage{multicol}

\def\BibTeX{{\rm B\kern-.05em{\sc i\kern-.025em b}\kern-.08em
    T\kern-.1667em\lower.7ex\hbox{E}\kern-.125emX}}

\begin{document}
\pagestyle{fancy}
\fancyhead[C]{This paper appears in IEEE International Conference on Computer Communications (INFOCOM) PhD symposium, 2023.}

\title{Semantic communications, semantic edge computing, and semantic caching\\[-9pt]}

\author{\IEEEauthorblockN{Wenhan Yu, Jun Zhao\\Nanyang Technological University, Singapore\\wenhan002@e.ntu.edu.sg, junzhao@ntu.edu.sg\vspace{-13pt}}} 

\maketitle
\thispagestyle{fancy}
\begin{abstract}
The increasing popularity of applications like the Metaverse has led to the exploration of new, more effective ways of communication. Semantic communication, which focuses on the meaning behind transmitted information, represents a departure from traditional communication paradigms. As mobile devices become increasingly prevalent, it is important to explore the potential of edge computing to aid the semantic encoding/decoding process, which requires significant computing power and storage capabilities. However, establishing knowledge bases (KBs) for domain-oriented communication can be time-consuming. To address this challenge, this paper proposes a semantic caching model in edge computing system that caches domain-specialized general models and user-specific individual models. This approach has the potential to reduce the time and resources required to establish individual KBs while accurately capturing the semantics behind users' messages, ultimately leading to more efficient and accessible semantic communication.

\end{abstract}


\section{Introduction}

\begin{figure*}[t] 
\centering
\setlength{\abovecaptionskip}{-0.1cm}
\includegraphics[width=0.9\linewidth]{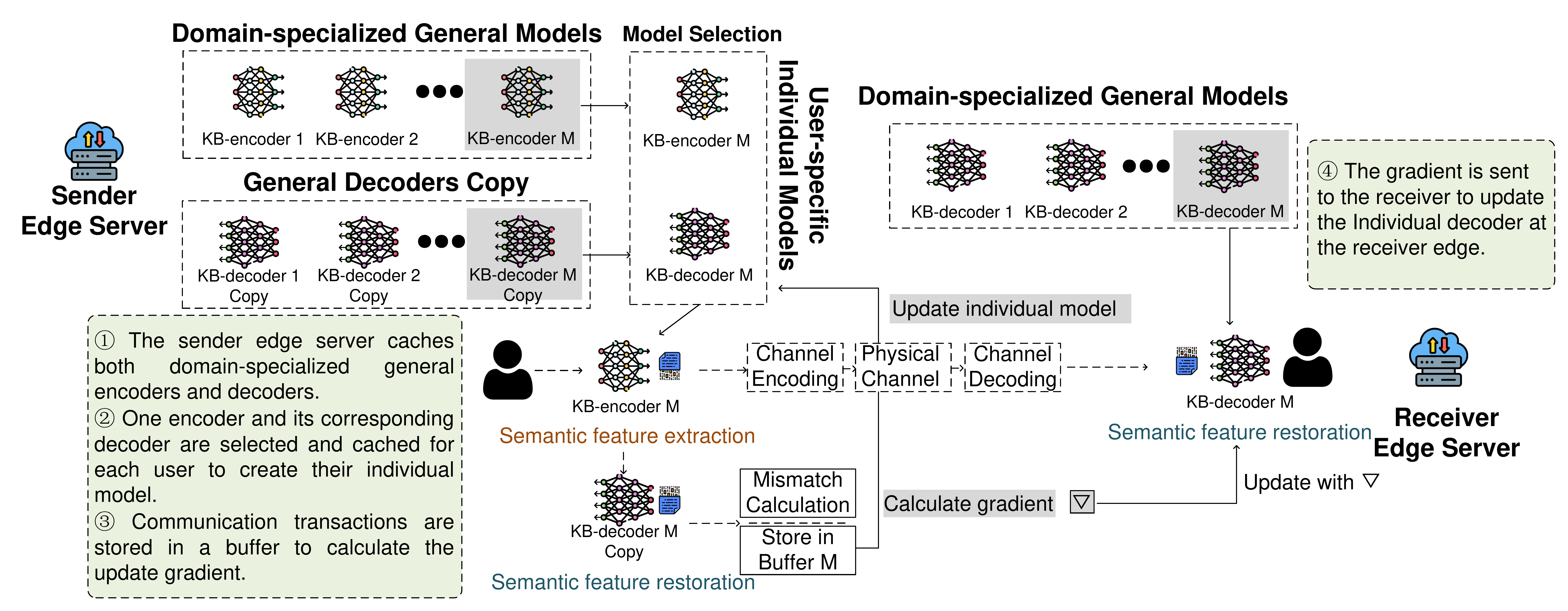}
\caption{Semantic Edge Computing and Caching Model.}\vspace{-0.6cm}
\label{fig:alg}
\end{figure*}

As wireless communication technology continues to evolve, emerging applications such as the Metaverse and ultra-reliable, low-latency communications have become increasingly popular, prompting people to seek more effective and innovative ways of communication. Semantic communication is a departure from traditional communication paradigms, which transmit data bit by bit, by focusing on the meaning behind the transmitted information. In a semantic communication system, the workflow typically includes the following steps: Semantic encoding, Channel encoding, Physical channel, Channel decoding, and Semantic decoding~\cite{semanticmag}. Unlike traditional communication systems, semantic communication systems rely on Semantic encoding, which extracts semantic features from the source, and Semantic decoding, which restores the message with the features. Deep learning techniques are commonly used to support semantic encoding and decoding.

Given the growing prevalence of mobile devices in our daily lives, it is essential to explore the potential of edge computing to aid the semantic encoding/decoding process, as semantic communication requires a certain level of computing power and storage capabilities. Semantic feature extraction and restoration are the most important parts of the process, which are carried out using specified knowledge bases (KBs) from both ends. Typically, these KBs are deep-learning models that self-learn over time. Communication between different users is specialized and usually domain-oriented, making the establishment of the KBs time-consuming.

To address this challenge, we propose a semantic caching model in the semantic edge computing system. This model aims to assist the semantic communication process by caching the domain-specialized general models, and user-specific individual models derived from the general models for each user. 

\section{Semantic caching in semantic edge computing}
\subsection{Domain-Specialized General Model}
Using only general models for all users can lead to severe mismatches between senders and receivers. For example, the word "bus" may refer to a transportation vehicle in our daily life, but it may mean a high-speed internal connection in computer architecture. Additionally, for an image with multiple elements in different domains, the focus may lie on different things. Therefore, it is more reasonable to consider multiple models specialized by data from different domains. To ensure effective communication between users on different edge servers, we assume that each sender edge server $i$ caches multiple well-pretrained general KB-encoders specialized for different major domains such as IT, medical, news, and entertainment, denoted by $\{e_i^1, e_i^2,\ldots,e_i^N\}$, where $\mathcal{M}=\{1,2,\ldots,M\}$ represents the different domains. During communication, we select a KB-encoder $e_i^m$ ($m\in\mathcal{M}$ as the domain) based on the domain of the messages being transmitted and use it to extract semantic features. We then transmit these features to the receiver edge server $j$, which uses the corresponding KB-decoder $d_j^m$ to restore the features to complete messages.

\subsection{User-Specific Individual Model}
Besides, the domain-specialized general model for semantic communication without considering the user's specific language patterns and communication styles can also lead to problems. The general model may not accurately capture the nuances and context-specific language usage of individual users. For example, different people may use the same word or phrase to mean different things depending on their cultural background, age, or personal experiences. A general model may not be able to capture these subtle differences, leading to misunderstandings and misinterpretations. Therefore, training a user-specific model is important for semantic communication because it allows for more personalized and effective communication, improving the accuracy and efficiency of communication between machines and humans.

\subsection{Decoder Copy on Sender Edge}
Nevertheless, training a user-specific KB model for each user, such as $u_1$, to alleviate semantic errors creates a challenge in identifying semantic mismatches between encoders and decoders across different edge servers. This is because calculating the mismatches requires both input and output data, which are located on different servers. Sending the output back to the sender would defeat the purpose of the semantic communication system, which aims to decrease the transmitted data sizes. To address this challenge, we cache general decoders $\{d_i^1, d_i^2,\ldots,d_i^N\}$ at both the sender edge server $i$ and receiver edge server $j$, which means $d_j^m=d_i^m (\forall m\in\mathcal{M})$. After each communication, the mismatch can be calculated on the server, and we store this information in the domain buffer $b_m$ for data collection.

\subsection{Update Process}
We use $e_{u_1}^m, d_{u_1}^m$ at sender server $i$ as the user-specialized model evolved from the general models $e_i^m, d_i^m$, and they start to be trained together after enough collected data at $b_m$. Then, the gradient of decoder $\nabla d_{u_1}^m$ will be transmitted to the receiver $j$ to synchronize the $d_{u_2}^m$, which is similar to the update process in traditional Federated Learning (FL)~\cite{FLsurvey}. Note that the general models remain the same during all time, and the user-domain-specialized model will be cached separately.

\section{Research Direction}
Under such a semantic communication system, there are several directions we can work on.

\subsection{Model Selection}
One promising research direction is exploring the selection of a general model based on the user's communication topic. While a traditional classification neural network can be used to determine which domain the message belongs to, it may not take into account the context of the message. As context is often critical in selecting the appropriate model, deep reinforcement learning or LSTM-based classification networks can be utilized to better evaluate the best way of selecting the model. By leveraging these advanced techniques, the selection process can be optimized based on the specific communication context, taking into account not only the content of the message, but also the user's preferences and habits. This personalized approach can greatly enhance the effectiveness and efficiency of the semantic communication system, leading to more accurate and relevant responses for users.

\subsection{Encoding and Decoding Models}
The encoding and decoding models play a critical role in a semantic communication system. Given the diverse nature of message types, including text, image, video, and audio, it is crucial to consider multimodality when designing these models. A well-designed model should be capable of handling different data types and extract relevant semantic features accurately. In addition, it should be able to efficiently encode and decode messages to enable effective communication. Therefore, researchers have been exploring various deep learning techniques to develop effective encoding and decoding models that can handle multimodal data efficiently. Some promising approaches include convolutional neural networks (CNNs), recurrent neural networks (RNNs), and transformer-based models.

\subsection{Communication Optimization}
In addition to the aforementioned aspects, the proposed semantic communication system also offers the opportunity to study and optimize various communication problems. For instance, issues such as signal interference, transmission errors, and network congestion can be addressed and mitigated through effective channel encoding and decoding techniques. Furthermore, the use of different deep learning algorithms and edge computing technologies can be testified to improve the overall system performance by enhancing data processing speed and reducing latency. Moreover, the system's ability to extract and utilize semantic features can also be accelerated to give better user experience and satisfaction. Additionally, other communication problems such as security, privacy, and reliability can also be studied and addressed in this system.

\renewcommand{\refname}{References}


\end{document}